\input{aipcheck}

\documentclass[
    ,final            
  ]
  {aipproc}

\layoutstyle{6x9}

\begin{document}

\title{The Living with a Red Dwarf Program: Observing the Decline in dM Star
FUV Emissions With Age}

\keywords {stars: red dwarfs, dM stars, rotation, age; stars: individual
(CD-64 1208, AU Mic, HIP 107345, HIP 31878, AD Leo, EV Lac, Proxima Cen,
IL Aqr, SZ UMa); FUV; FUSE
\vspace{-5mm}}
\vspace{-5mm}

\author{Scott G. Engle}{
  address={Villanova University, Department of Astronomy \& Astrophysics,
           800 E. Lancaster Ave, Villanova, PA 19085, USA}
  ,altaddress={James Cook University, Centre for Astronomy, Townsville
               QLD 4811, Australia}
}

\author{Edward F. Guinan}{
  address={Villanova University, Department of Astronomy \& Astrophysics,
           800 E. Lancaster Ave, Villanova, PA 19085, USA}
}

\author{Trisha Mizusawa}{
  address={Villanova University, Department of Astronomy \& Astrophysics,
           800 E. Lancaster Ave, Villanova, PA 19085, USA}
}

\begin{abstract}
Red Dwarf (dM) stars are overwhelmingly the most numerous stars in our
Galaxy. These cool, faint and low mass stars make up $>$ 80\% of all stars.
Also dM stars have extremely long life times ($>$50-100 Gyr). Determining the
number of red dwarfs with planets and assessing planetary habitability (a
planet's potential to develop and sustain life) is critically important
because such studies would indicate how common life is in the universe.
Our program -- ``Living with a Red Dwarf'' -- addresses these questions by
investigating the long-term nuclear evolution and the coronal and
chromospheric properties of red dwarf stars
with widely different ages ($\sim$50 Myr -- 12 Gyr). One major focus of the
program is to study the magnetic-dynamo generated coronal and chromospheric
X-ray-FUV/UV emissions and flare properties of a sample of dM0--5 stars.
Observations carried out by FUSE of a number of young to old dM stars
provide important data for understanding transition region heating in these
stars with deep convective zones as well as providing measures of FUV
irradiances. Also studied are the effects of X-ray--FUV emissions on possible
hosted planets and impacts of this radiation on their habitability. Using
these data we are constructing irradiance tables (X--UV irradiances) that
can be used to model the effects of XUV radiation on planetary atmospheres
and possible life on planetary surfaces. The initial results of this program
are discussed.
\end{abstract}

\maketitle


\vspace{-3mm}
\section{Introduction \& Background}
\vspace{-1mm}

The ``Living with a Red Dwarf'' program aims to understand the magnetic
activity, dynamo structure and plasma physics (along with determining the
X-ray--FUV--UV [XUV] spectral irradiances) of dM stars with widely different
rotations, ages and correspondingly different levels of magnetic activity.
For a more in-depth summary of the program and some preliminary results,
see Guinan \& Engle (2009).
This program is an extension of the ongoing ``Sun in Time'' program which
primarily targets main-sequence G-stars (Guinan et al. 2003). Another
important aspect of the Living with a Red Dwarf program is the determination
of rotation periods for dM stars of reliably known ages, in order to form
a well-populated Age-Rotation relationship. This is done through available
survey photometry (such as the invaluable All Sky Automated Survey -- ASAS
-- Pojmanski (2001))
in addition to continued multi-band photometry we have carried out for the
past $\sim$5 years. When accomplished, the rotation and age data can be
combined with our XUV irradiances and emissions to delineate dM
star magnetic evolution over time.

The FUSE spectral region contains important chromospheric and transition
region diagnostic emission lines that can be used to characterize energy
transfer and stellar atmospheric structure while the ratio of C {\sc iii}
1176/977 fluxes contains valuable information about the electron pressure. 
This spectral region is also important for gauging the FUV emissions of
dM stars (mostly contributed by the H {\sc i} Lyman series, C {\sc iii}
and O {\sc vi} emissions). This is critical to assess the photochemical
and photoionization evolution of planetary atmospheres and ionospheres
and the possible development of life on extrasolar planets (Kasting 1997;
Guinan \& Ribas 2002; Kasting \& Catling 2003; Guinan et al. 2003; Ribas
et al. 2005). Table 1 lists the dM stars observed by FUSE and some of 
their physical properties most relevant to the program. Stellar ages have
been assigned as follows: CD-64 1208 -- $\beta$ Pic Association member;
AU Mic -- $\beta$ Pic Association; HIP 107345 -- Tucana-Horologium
Association; HIP 31878 -- AB Dor Moving Group, Subgroup B4;  AD Leo --
Rotation-Age Relation; EV Lac -- UMa Group; Proxima Cen -- $\alpha$ Cen
System isochronal fits; IL Aqr -- Rotation-Age; SZ UMa -- Old Disk Space
Motions (memberships from L\'opez-Santiago et al. (2006) \& Montes et al.
(2001)). Our program had an approved proposal for FUSE observations to
begin filling in the large age gap for dM stars of $\sim$1--6 Gyr but, 
unfortunately, the FUSE satellite suffered its fatal malfunction before
the observations could be carried out.

 
\begin{figure}
  \includegraphics[height=.8\textheight]{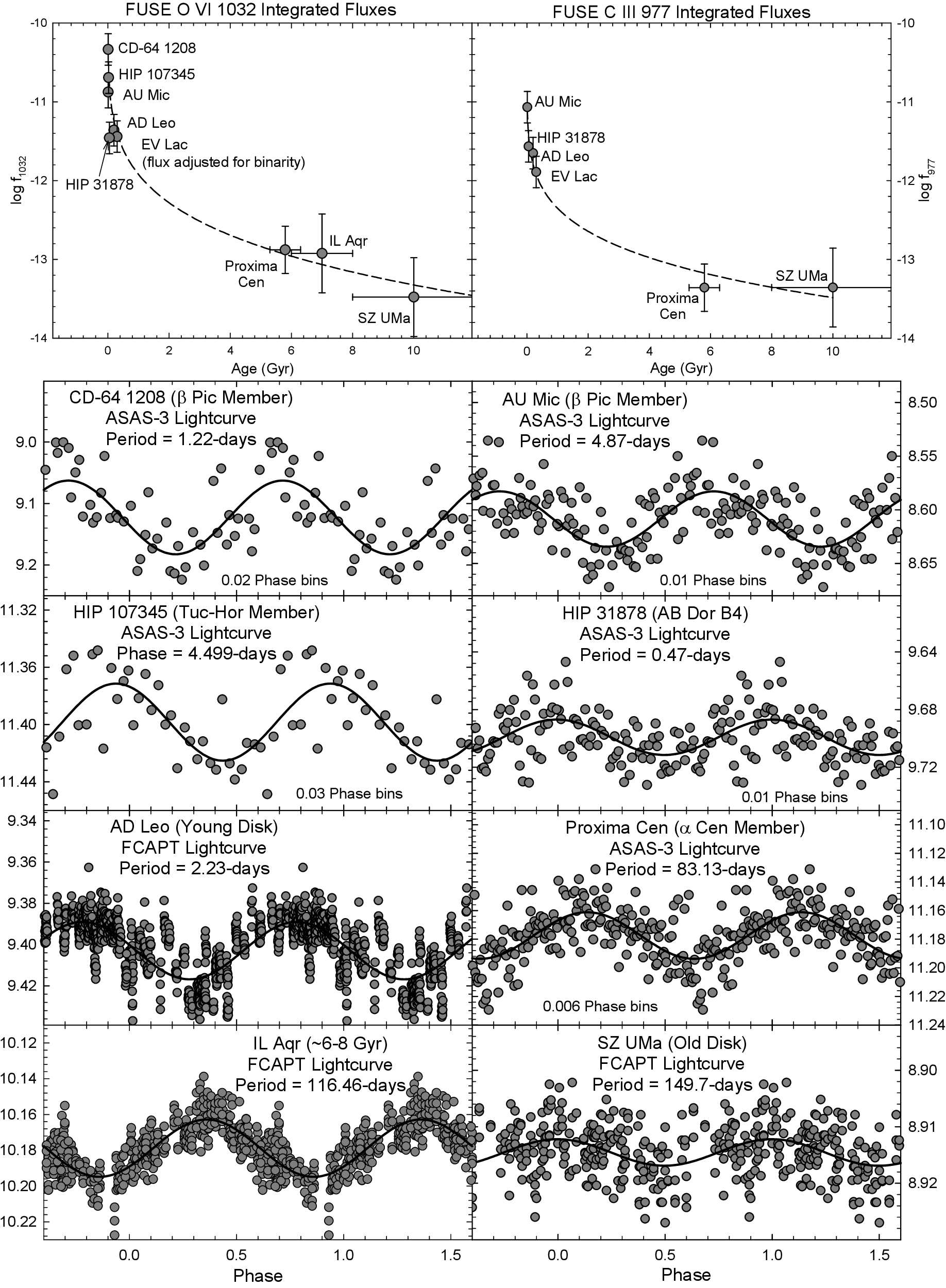}
  \caption{
{{\bf Top --} The FUV (O {\sc vi} [left] and C {\sc iii}) [right] integrated
emission line fluxes from FUSE observations are plotted, indicative of 
transition region and chromospheric activity, respectively. One can easily see
the diminishing emissions of these two atmospheric layers as the stars age.
{\bf Bottom --} The rotation V-band light curves for 8 of the FUSE observed
dM stars are shown, making use of either ASAS-3 survey photometry,
or FCAPT pointed photometry that we continue to carry out. Photometrically 
determined rotation periods can be much more accurate than spectroscopically
determined ones, especially for longer rotation periods. Pertinent information
for the target star and lightcurve are given in each plot.}}
\end{figure}


\begin{table}{\small}
\begin{tabular}{llllll}
\hline
{\bf Star Name} & {\bf FUSE Name} & {\bf V-mag} & {\bf Spec. Type}
& {\bf Age} & {\bf Rotation} \\
\hline
CD-64 1208  & CD-64D1208 & 9.2   & M0          & 0.010 ($\beta$ Pic)& 1.22 \\
AU Mic      & HD197481   & 8.61  & M0Ve        & 0.010 ($\beta$ Pic)& 4.86 \\
HIP 107345  & HIP107345  & 11.72 & M1          & 0.025 (Tuc-Hor)    & 4.5  \\
HIP 31878   & CD-61D1439 & 9.70  & M1V         & 0.050 (AB Dor B4)  & 0.47 \\
AD Leo      & BD+20D2465 & 9.43  & M4.5Ve      & 0.200 (Rotation)   & 2.23 \\
EV Lac      & BD+43D4035 & 10.09 & dM4.5+dM4.5 & 0.300 (UMa Group)  & 4.38 \\
Proxima Cen & GJ551      & 11.05 & dM5e        & $\sim$5.8$\pm$0.4 ($\alpha$ Cen System) & 83.2 \\
IL Aqr      & GLIESE876  & 10.17 & dM5         & $\sim$6--8 (Rotation) & 116.46\\
SZ UMa      & SZUMA      & 9.32  & M1V         & 8--12 (Old Disk)   & 149.7\\
\end{tabular}
\caption{dM Stars in the FUSE Archive}
\label{tab:a}
\end{table}

\vspace{-3mm}
\section{Preliminary Results of the Program}
\vspace{-1mm}

The top two plots of Fig. 1 illustrate the declining transition region
(O {\sc vi} 1032\AA) and chromospheric (C {\sc iii} 977\AA) emissions of
dM stars with age, as observed
by the FUSE satellite. Plotted are the integrated fluxes of the respective
emission lines, all adjusted to the distance of Proxima Cen. Stars with no
C {\sc iii} emissions plotted did not have sufficient night-time data to
permit a measurement of the line. Note how the strongest emissions are from
the three Pre-Main Sequence (PMS) stars -- CD-64 1208, AU Mic \& HIP 107345.
This is most likely {\em not} an indicator of absolute emission strength,
but a result of the increased surface area of the PMS stars, which have not
yet fully contracted on to the Main Sequence. The two plots are set to the
same scales, so one can easily see the nearly identical behaviors of
the chromosphere \& transition region. 

The lower plots of Fig. 1 show the light curves for 8 of the FUSE targets
(EV Lac has not yet
been observed by us and is too far north for ASAS-3 photometry, but has a
previously determined photometric rotation period of 4.376-days -- Contadakis
(1995)). The three ``slow rotators'' in the FUSE database are of particular
interest. 

{\bf Proxima Cen} is a member of the $\alpha$
Cen system. The rather accurate age of the star ($\sim$5.8 Gyr) comes from
extensive isochronal studies of its companion star, $\alpha$ Cen A. Such
an accurate and ``old'' age for a dM star is singular to Proxima Cen and,
accordingly, it has always been a star of great importance in our studies
of the relationship between rotation, activity and age. 

{\bf IL Aqr} is presently known to host three planets, two of which are ``Super
Earths.'' We have observed IL Aqr with the Four College Automatic
Photoelectric Telescope (FCAPT), at Fairborn Observatory, since 2004. Until
recently, our best fitting rotation period was in close agreement with that
of Rivera et al. (2005). This past year, however, a better-covered
minimum was observed and our best fitting period was lengthened to
its current value of $\sim$116.46-days. This indicates an age even older than
that of Proxima Cen for a star with three currently known planets, and is
rather exciting.

We successfully proposed for a FUSE observation of {\bf SZ UMa} because its
{\em UVW} space motions
(133, -8, 0 -- Delfosse et al. 1998) imply Old Disk--Halo membership and an
age of $\sim$8--12 Gyr. Its rotation period of $\sim$149.7-days (determined
from FCAPT photometry) is one of the longest we have found to date for a dM
star.

\vspace{-3mm}
\section{Discussion}
\vspace{-1mm}

The photometrically determined rotation periods of all 9 dM stars in the
FUSE database have been presented. Eight of these rotation periods have been
determined by us from recently obtained survey (ASAS-3) or pointed (FCAPT)
photometry. The ages of the dM stars have also been given along with their
integrated C {\sc iii} 977\AA~ and O {\sc vi} 1032\AA~ emission line fluxes,
as observed by the FUSE satellite. This has permitted a study of the decline
in dM star chromospheric and transition region emissions as they age. The
data will be of use in understanding the FUV environments of dM stars, their
magnetic activity over time, and also the possible habitability of planets
hosted by them. 


\vspace{-3mm}
\begin{theacknowledgments}
\vspace{-2mm}
This research is supported by grants from NASA/FUSE (NNX06AD38G), and NSF
(AST-0507542 \& AST-0507536) which we gratefully acknowledge. We also
gratefully acknowledge the invaluable and continued contributions of
Dr. Grzegorz Pojmanski.
 
\end{theacknowledgments}



\bibliographystyle{aipproc}   

\bibliography{sample}

\IfFileExists{\jobname.bbl}{}
 {\typeout{}
  \typeout{******************************************}
  \typeout{** Please run "bibtex \jobname" to optain}
  \typeout{** the bibliography and then re-run LaTeX}
  \typeout{** twice to fix the references!}
  \typeout{******************************************}
  \typeout{}
 }

\end{document}